# Spin-Orbital Separation in the quasi 1D Mott-insulator Sr$_2$CuO$_3$

J. Schlappa[1,2*], K. Wohlfeld[3], K. J. Zhou[1], M. Mourigal[4], M. W. Haverkort[5], V. N. Strocov[1], L. Hozoi[3], C. Monney[1], S. Nishimoto[3], S. Singh[6#], A. Revcolevschi[6], J.-S. Caux[7], L. Patthey[1], H. M. Rønnow[4], J. van den Brink[3], and T. Schmitt[1*]

[1]Paul Scherrer Institut, Swiss Light Source, CH-5232 Villigen PSI, Switzerland

[2]Institut Methoden und Instrumentierung der Forschung mit Synchrotronstrahlung G-I2, Helmholtz-Zentrum Berlin für Materialien und Energie GmbH, D-12489 Berlin, Germany

[3]Institute for Theoretical Solid State Physics, IFW Dresden, Helmholtzstrasse 20, 01069 Dresden, Germany

[4]Ecole Polytechnique Fédérale de Lausanne (EPFL), CH-1015 Lausanne, Switzerland

[5]Max Planck Institute for Solid State Research, D-70569 Stuttgart, Germany

[6]ICMMO - UMR 8182 - Bât. 410, Université Paris-Sud 11, 91405 Orsay Cedex, France

[7]Institute for Theoretical Physics, University of Amsterdam, Science Park 904, Postbus 94485, 1090 GL Amsterdam, The Netherlands

Present address:

[#]Indian Institute of Science Education and Research, 900 NCL Innovation Park, Pashan – 411008 Pune, India

[*]**Corresponding authors: justine.schlappa@helmholtz-berlin.de & thorsten.schmitt@psi.ch**

**As an elementary particle the electron carries spin $\hbar/2$ and charge $e$. When binding to the atomic nucleus it also acquires an angular momentum quantum number corresponding to the quantized atomic orbital it occupies (e.g., *s*, *p* or *d*). Even if electrons in solids form bands and delocalize from the nuclei, in Mott insulators they retain their three fundamental quantum numbers: spin, charge and orbital[1]. The hallmark of one-dimensional (1D) physics is a breaking up of the elementary electron into its separate degrees of freedom[2]. The separation of the electron into independent quasi-particles that carry either spin (spinons) or charge (holons) was first observed fifteen years ago[3]. Using Resonant Inelastic X-ray Scattering on the 1D Mott-insulator $Sr_2CuO_3$ we now observe also the orbital degree of freedom separating. We resolve an orbiton liberating itself from spinons and propagating through the lattice as a distinct quasi-particle with a substantial dispersion of ~0.2 eV.**

It was pointed out in the 1970's by Kugel and Khomskii[1] that in a solid not only the charge and spin of electrons can order – leading to magnetism – but also the electrons' orbital degree of freedom. This observation sparked a field that has blossomed since. While a physical electron knits together spin, charge and orbital, theoretically an electron can in principle be considered as a bound-state of the three independent fundamental quasi-particles: a spinon carrying its spin, a chargon (or holon) carrying its charge, and an orbiton carrying its orbital degree of freedom.

A remarkable and very fundamental property of 1D systems is that electronic excitations break up into deconfined spinons and holons, carrying separately spin and charge, respectively. This was predicted decades ago[2] and confirmed in the mid 1990's by angular-resolved photo-emission spectroscopy (ARPES) experiments[3-5]. The spin-charge separation is an example of particle fractionalization, the phenomenon in which quantum numbers of the quasi-particles are not multiples of those of the elementary electron, but fractions. This effect is one of the most spectacular manifestations of collective quantum physics of interacting particles and is a profound concept that has found its way into a number of theories, e.g., in relation to the phenomenon of high temperature superconductivity in cuprates[6-7].

To search for the further fractionalization of the electron, we consider the excitation of a copper orbital degree of freedom in the antiferromagnetic (AF) spin chain compound $Sr_2CuO_3$. The spin-orbital separation process that we are looking for is analogous to the spin-charge separation mechanism (see Fig. 1b). The latter occurs for instance when an electron is annihilated, removing a single spin and leaving behind a hole in the AF chain. This hole can start propagating freely only after exciting one spinon (a domain wall in the AF chain). Subsequently also the spinon can delocalize and separate itself completely from the holon. When instead of creating a hole, as one typically does in a photoemission experiment, an electron is excited from one copper 3$d$ orbital to another, the phenomenon of spin-orbital separation can in principle occur (see Fig. 1a). The orbiton created in this manner may also deconfine after exciting a spinon, thus splitting up the electron in its orbital and spin degree of freedom[8].

Here we use high-resolution Resonant Inelastic X-ray Scattering (RIXS) to search experimentally for spin-orbital separation in the quasi-1D cuprate $Sr_2CuO_3$ [for material details see Sec. 1 of supplementary information]. We observe deconfinement of the spinon and orbiton, when making an orbital excitation from the ground state copper 3$d$ $x^2$-$y^2$ orbital to an excited copper 3$d$ $xy$ or $xz$ orbital (Fig. 1c-e)[*]. We measure an orbiton dispersion that is almost as large as the dispersion of the two-spinon continuum at low energies. Similarly as for spin-charge separation[3-5], the orbiton dispersion has π periodicity (see also Fig. 4b below), which is the smoking gun for the presence of an orbiton liberated from the spinon.

We measured the orbital excitations of $Sr_2CuO_3$ using RIXS at the $L_3$ edge of the Cu ion. RIXS is a second order scattering technique and can excite transitions between the Cu 3$d$ states of different symmetry (orbital excitations), due to the involvement of two subsequent electric dipole transitions[9-10] [see Sec. 3 of supplementary information].

---

[*] For simplicity we will from here on use the so-called 'hole' language: while nominally there are 9 electrons in the 3$d$ orbitals of the $Cu^{2+}$ ion in $Sr_2CuO_3$, using the 'electron-hole' transformation, the problem can be mapped onto an effective system with one particle occupying a single 3$d$ orbital, see Sec. 2 of the supplementary information.

With the unique capability of RIXS to probe also spin excitations[11-13] and to vary the photon-momentum transfer, the dispersion of orbital and spin excitations can be mapped out across the first Brillouin Zone (BZ)[11-16]. The experiments were carried out at the ADRESS beamline of the Swiss Light Source at the Paul Scherrer Institut[17-18].

For fixed momentum transfer along the chains, $q$, peaks in the RIXS spectrum situated at constant energy transfer reveal the presence of charge-neutral elementary excitations and are clearly present in the RIXS intensity map of $Sr_2CuO_3$ across the Cu $L_3$ edge in Fig. 2a. The spectrum with incident energy precisely tuned at the resonance maximum of the absorption spectrum is shown in Fig. 2b. In both graphs the excitations of the spin, orbital and charge degrees of freedom are indicated. The momentum dependence and in particular the dispersion of the spin and orbital excitations, see Fig. 1c (as well as fig. S2a in supplementary information), is indicative for their collective nature.

While up to ca. 0.8 eV energy transfer purely magnetic excitations are present, the spectrum between ~1.5 and ~3.5 eV corresponds to excitations from the Cu 3d $x^2$-$y^2$ ground state to orbitals of $xy$, $xz/yz$ and $3z^2$–$r^2$ symmetry (see Fig. 1d and 1e). These peaks correspond to orbital excitations (called also $d$-$d$ excitations), and not, e.g., to charge transfer excitations, the intensity of which extends up to ca. 6 eV and is at least an order of magnitude lower in $L$ edge RIXS, cf. Ref. 10. The orbital assignment of these excitations was unambiguously verified by comparing their energy at $q$=0 with *ab initio* quantum chemistry cluster calculations[19] [see Sec. 5 of supplementary information for detailed results].

Zooming into the magnetic part of Fig. 1c between 0 – 0.8 eV in energy transfer reveals strongly dispersing spin excitations, where the lower boundary has period $\pi$ and the continuum (upper boundary) period $2\pi$ (Fig. 3a). These RIXS data agree very well with recent inelastic neutron scattering (INS) studies on $Sr_2CuO_3$[20]. The simultaneous presence in the spectrum of a lower edge with period $\pi$ and an upper one with period $2\pi$ indicates directly that in the spin chain the magnetic excitations with spin 1 break up and fractionalize into two-spinon (and higher order) excitations that make up a continuum[20], with each spinon carrying a spin ½. These spectra confirm that RIXS for magnetic excitations probes the well-known spin dynamical structure factor as theoretically predicted[11-13], in agreement with recent studies on TiOCl[21]. The excellent statistics of the data further allows for a direct comparison of the RIXS line shapes with the exact two- and four-spinon dynamical structure factor of the spin-½ Heisenberg chain. In Fig. 3b the fit for three selected momentum-transfer values is displayed using the exact two- plus four-spinon dynamical structure factor $S(q,\omega)$ in the representation of Caux and Hagemans[22]. The obtained exchange coupling $J$=249 meV is in perfect agreement with the value obtained from INS data[20].

Having identified unambiguously the fractionalized spinon excitations in the low-energy sector, we now concentrate on the orbital excitations spectrum in Fig. 4. We see that these are strongly momentum dependent and exhibit a distinct dispersion that has never been observed before. This proves that the orbital excitations observed here are of collective nature. The $xz$ excitation has the largest dispersion, of approx. 0.2 eV, with a spectrum containing two peculiar components: a lower branch dispersing with periodicity $\pi$ and an incoherent spectrum on top of it with a double-oval shape. This spectrum is strikingly similar to seemingly unrelated ARPES spectra of 1D cuprates which evidence spin-charge separation [see e.g. Fig. 3 of Ref. 3]. This is a first indication that the observed orbital dispersion is related to an analogous separation of degrees of freedom.

To test this conjecture we have derived the microscopic model that describes the spin-orbital interactions in $Sr_2CuO_3$ [see Methods]. The proper low-energy Kugel-Khomskii Hamiltonian was obtained[8,23] from the charge transfer model of $Sr_2CuO_3$ in Ref. 24. The crucial part of the Hamiltonian, responsible for the $xz$ orbital propagation, is

$$H = -J_O \sum_{js}(c^+_{js}c_{j+1,s} + h.c.) + J\sum_j \vec{S}_j \vec{S}_{j+1} + E_O \sum_j (1-n_j), \quad (1)$$

where $J$ ($J_O$) is the spin (orbital) exchange constant, both of which are fixed by the charge transfer model[24]. $E_O$ is the $xz$ on-site energy. The first term in the Hamiltonian promotes the propagation of orbital excitations through the lattice. The second term represents the usual Heisenberg interaction between spins, which vanishes on bonds where an orbital excitation is present [see Methods for more details].

From purely a mathematical standpoint equation (1) is identical to the 1D $t$-$J$ model, with the orbital superexchange $J_O$ taking the place of $t$, the hole hopping amplitude in the $t$-$J$ model. This implies that the propagation of a single orbital excitation in the $J_O$-$J$ model above is equivalent to the propagation of a single hole in the 1D $t$-$J$ Hamiltonian. Since in the 1D $t$-$J$ model a hole breaks up into a free spinon and holon, in the $J_O$-$J$ model the orbital excitation will separate into a free spinon and orbiton.

Before calculating RIXS spectral functions and quantifying the separation of spin and orbital degrees of freedom in $Sr_2CuO_3$, it is instructive to consider the overall features of the momentum dependent RIXS spectra within a slave-boson picture in terms of an orbital-spin separation Ansatz (OSSA). This Ansatz is in spirit equivalent to the mean-field slave-boson picture for charge-spin separation[25-26], which stipulates that the hole spectral function is given by the convolution of the spectral function of a free holon and a free spinon. Analogously, in the OSSA free orbitons and spinons, created by the operators $o_i^+$ and $s_{i,\sigma}^+$, build the particle creation operator $c_{i,\sigma}^+ = s_{i,\sigma}^+ o_i$. Applying this formalism to the Hamiltonian in Eq. 1, the energy of noninteracting orbitons and spinons with momentum $k$ becomes $\varepsilon_O(k) = E_O + 2J_O \cos(k)$ and $\varepsilon_S(k) = -J \cdot \cos(k)$, respectively. In the OSSA the spectral function for the orbital excitation produced in RIXS, $c_{q,\sigma}^+ = \sum_k o_{q-k} s_{\sigma,k}^+$ (where **q** is here the momentum transfer along the $CuO_3$ chain direction), is determined by a convolution of these two dispersions. The resulting excitation continuum is indicated in Fig. 4b and has three defining features[25-26]: 1. a prominent lower edge with energy $E(q) = E_O - 2J_O |\sin(q)|$ that is entirely determined by the dispersion of the orbiton, 2. a distinctive upper orbiton edge at $E(q) = E_O + 2J_O |\sin(q)|$ and 3. a broader top of the orbiton-spinon continuum at $F(q) = E_O + \sqrt{J^2 + 4J_O^2 - 4J \cdot J_O \cos(q)}$. These dispersion curves are directly compared to our RIXS experiments in Fig. 4b. The measured spectral weight that piles up at the first two lower-energy features does so with periodicity $\pi$, which is a direct consequence of spin-orbital separation.

To quantify the spectral weights within the orbiton-spinon continuum, we have calculated for the $J_O$-$J$ Hamiltonian the orbital excitation Green's function by exactly diagonalizing the Hamiltonian on 28 lattice sites [as in the spin-charge separation studies, cf. Ref. 3., finite size effects are negligible (they are estimated at ca. 0.01 eV)]. From this the RIXS spectrum is calculated following Ref. 13, i.e., by expressing the RIXS amplitude as a product of the single-ion local RIXS effective operator and the orbital excitation Green's function [see Methods].

We find excellent agreement between theory and experiment (see Fig. 4): both the sine-like *xz* dispersion and the *xz* spinon-orbiton continuum (the "double-oval" incoherent spectrum) are present in the theory, which are, as already mentioned, the hallmarks of spin-orbital separation. The calculations also show that, in contrast to *xz*, the *xy* orbital has a rather small dispersion and that both the $3z^2$–$r^2$ and the *yz* orbital excitations are dispersionless, just as is observed experimentally. This is an independent merit of the model since no fitting of dispersions to experimental data is involved.

The large orbiton dispersion observed in this study is the key-feature that distinguishes $Sr_2CuO_3$ from other systems displaying orbital excitations[27-29]. For this the one-dimensional character of $Sr_2CuO_3$ is essential. In a system of higher dimensionality orbitons interact with magnetic excitations, which tend to slow them down and thus reduce their dispersion. In 1D orbitons can avoid these renormalization effects via spin-orbital separation.

**Methods summary**

We applied the technique of high-resolution Resonant Inelastic X-ray Scattering (RIXS) with the incident photon energy tuned to the $L_3$ edge ($2p_{3/2} \to 3d$ resonance) of the Cu ion (around 931 eV) and a total experimental resolution of 140 meV. The experiments were performed at the ADRESS beamline of the Swiss Light Source at the Paul Scherrer Institut[17-18]. The RIXS spectrometer was located at fixed scattering angle of either $\Psi=90°$ or $130°$ and was collecting signal within the solid angle of $(19.3\times3.3)$ mrad$^2$ (H$\times$V) (for a sketch of the experimental geometry see supplementary information, fig. S1).

$Sr_2CuO_3$ single-crystal samples were grown with the floating-zone method and freshly cleaved before the RIXS experiment. Surface normal (010) of the sample and the propagation direction of the chains (100) were oriented parallel to the scattering plane. The sample was cooled with a He-flow cryostat to 14 K during the measurements. Incident photons were linearly polarized in the scattering plane ($\pi$-orientation). The momentum transfer along the chains, $q$, was varied by changing the incidence angle in steps of $5\pm1°$.

Theoretical modelling was done by deriving a proper Kugel-Khomskii Hamiltonian from the charge transfer model and then writing it in a representation similar to the one used for the $t$-$J$ model. Using this model we calculated the spectral functions for the case of a single orbital excitation in the antiferromagnetic background by solving the Hamiltonian numerically on 28 sites. Such calculations were repeated separately for the two distinct dispersive orbital excitations, while for the nondispersive ones it was checked that indeed the parameters of the model did not allow for a dispersion larger than the experimental resolution. Finally, the RIXS cross section was calculated using the calculated spectral functions for the orbital excitations and multiplying it by the local RIXS form factors.

**Acknowledgments**

This work was performed at the ADRESS beamline using the SAXES instrument jointly built by Paul Scherrer Institut, Switzerland and Politecnico di Milano, Italy. We gratefully acknowledge support of the Swiss National Science Foundation and its NCCR MaNEP. K.W. acknowledges support from the Alexander von Humboldt foundation and fruitful discussions with M. Daghofer and S.-L. Drechsler. J.-S. C. acknowledges support from the Foundation for Fundamental Research on Matter (FOM) and from the Netherlands Organisation for Scientific Research (NWO). S.S. and A.R. acknowledge the support of the European contract NOVMAG. This research benefited from the RIXS collaboration supported by the Computational Materials Science Network (CMSN) program of the Division of Materials Science and Engineering, U.S. Department of Energy, Grant No. DE-SC0007091.

**Figure 1 | Spin-orbital separation process in an AF spin chain, emerging after exciting an orbital. a-b,** Sketch of spin-orbital separation **(a)** and spin-charge separation **(b)**, generated in a RIXS and ARPES process, respectively. **c,** RIXS intensity map of the dispersing spin- and orbital excitations in $Sr_2CuO_3$ *vs.* photon-momentum transfer along the chains and photon-energy transfer (for details see main text and Sec. 3 of supplementary information). **d**, Geometry of the $CuO_3$ chain with the ground state copper $3d$ $x^2$-$y^2$ orbitals in the middle of each plaquette and oxygen sites at the plaquette corners. **e,** Orbital symmetries of $x^2$-$y^2$ and excited $3d$ orbitals. In panels (c-e) 'hole' language is used (see Sec. 2 of supplementary information)

**Figure 2 | Energy dependence of elementary excitations in $Sr_2CuO_3$ observed with RIXS at the Cu $L_3$ resonance. a,** RIXS map of $Sr_2CuO_3$ *vs.* photon excitation energy (left axis) and energy transfer (bottom axis) on logarithmic intensity scale. The inserted black curve shows the total fluorescence yield (TFY) x-ray absorption spectrum. The dotted line marks the Cu $L_3$ resonance maximum energy. **b,** RIXS line spectrum

measured at the resonance maximum. All data was obtained at scattering angle $\Psi=90°$ and momentum transfer along the chain of $q = 0.189·2\pi/a$ (see Methods).

**Figure 3 | Dispersion of magnetic excitations: experimental data and simulation. a,** RIXS energy transfer spectra of the magnetic two-spinon continuum (zoom into the spin part from Fig. 1c). **b,** Fit (red line) of the experimental data (dots) with the two- plus four-spinon dynamical structure factor (green line)[22], convoluted with a Gaussian (shaded light grey region) to account for the total instrumental resolution of 140 meV. The low-energy peak around 50 meV denoted with a black line is elastic and phonon scattering. The momentum transfer $q$ of the selected spectra is indicated by arrows in (a).

**Figure 4 | Dispersion of orbital excitations: comparison between experiment and ab-initio calculations.** The orbital part from Fig. 1c (a, left) in comparison with theory, using the $J_O$-$J$ –model (c, right). The middle panel (b) shows the lower and upper edge of the orbiton dispersion (solid black and grey lines) and the upper edge of the spinon-orbiton continuum (dashed line), as calculated using the orbital-spin separation Ansatz. See also text and supplementary information for further details.

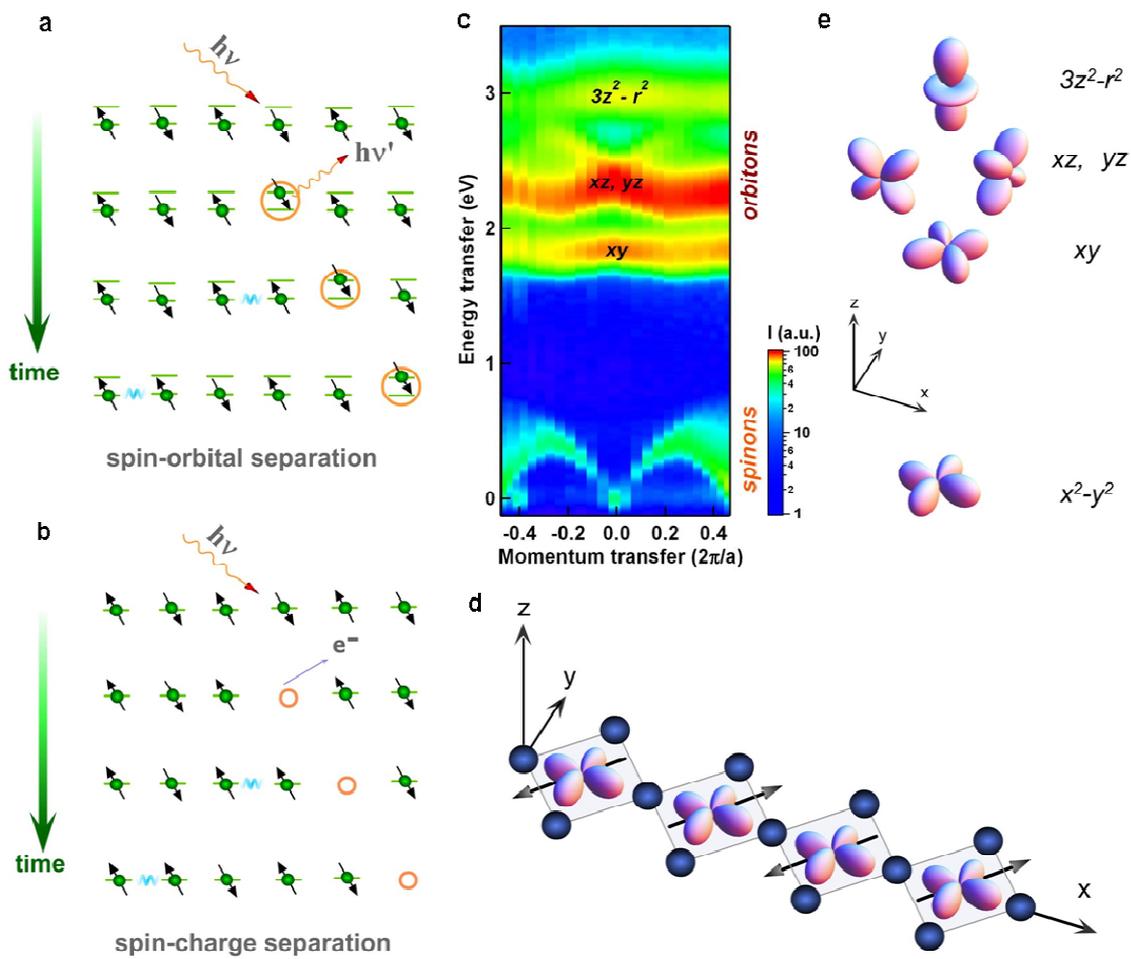

**Figure 1**

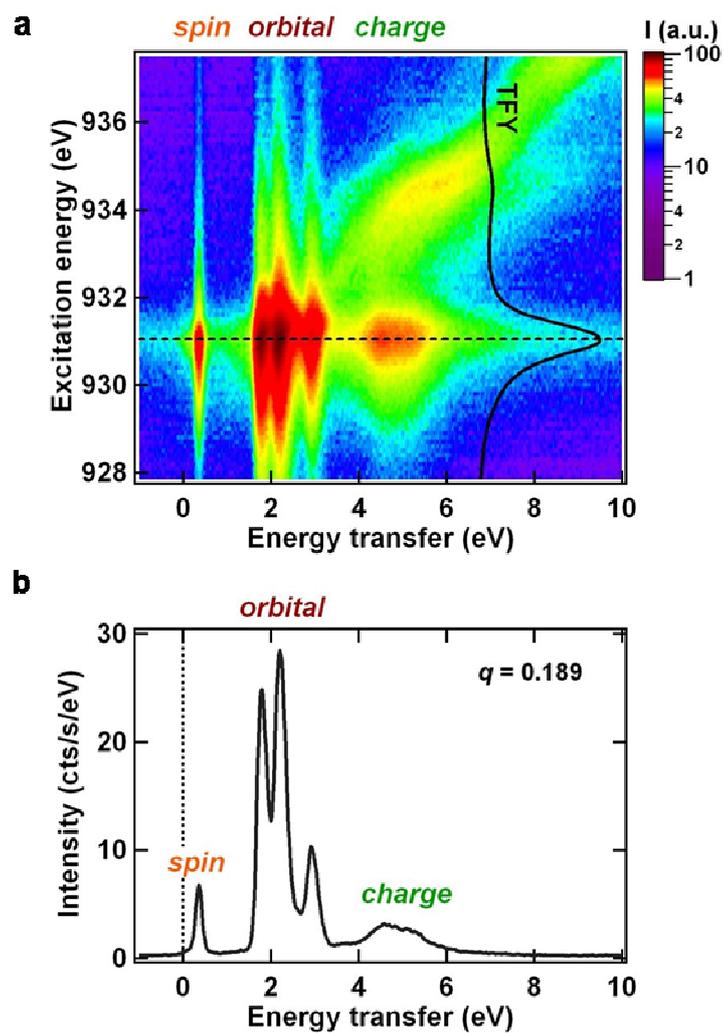

Figure 2

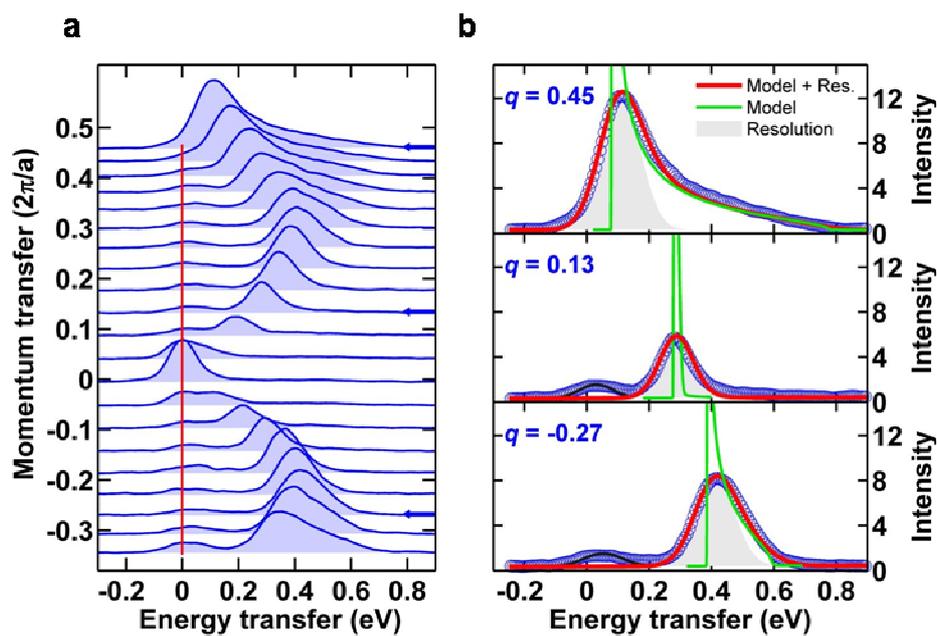

Figure 3

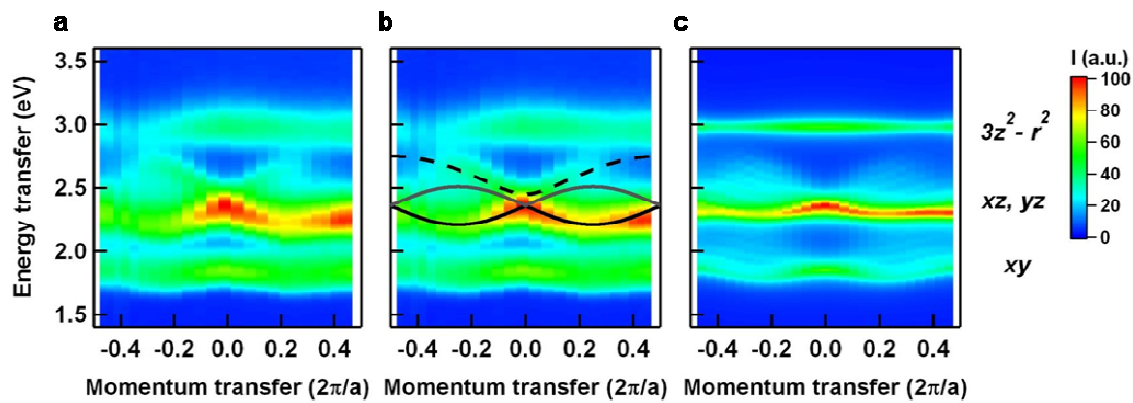

Figure 4

**Methods**

*A. Experiment*

Sr$_2$CuO$_3$ single-crystal samples were grown with the floating-zone method and freshly cleaved before the RIXS experiment. The samples were mounted such that the surface normal (010) and the chain propagation directions (100) were oriented parallel to the scattering plane [see Sec. 4 and fig. S1 of supplementary information]. The polarization vector of the incident light, $\varepsilon_{in}$, was parallel to the scattering plane ($\Pi$-orientation). This was yielding maximum cross section for Cu $2p^6 3d^9$ to Cu $2p^5 3d^{10}$ transition. The RIXS spectrometer was located at fixed scattering angle of either $\Psi = 90°$ or $130°$, collecting signal within the solid angle of $(19.3 \times 3.3)$ mrad$^2$ (H$\times$V). The momentum transfer along the chains, $q$, was varied by changing the incidence angle in steps of $5 \pm 1°$. This results in a polarisation dependence of the RIXS signal and a relative intensity variation of the orbital excitations with $q$. (For details on the polarization dependence of the orbital excitations see Sec. C2.) Fig. S2a shows the RIXS *line* spectra, which correspond to the intensity-map data in Fig. 1c and fig. S2b the variation of polarization of the incident photons with $q$. We estimate the systematic error in the momentum transfer to about 0.006 and 0.013 $2\pi/a$ at the edge and the center of BZ, respectively. The error in determining the energy zero of the energy-transfer scale is about 30 meV.

*B. Theory: spin excitations*

Due to the particular choice of the experimental geometry the RIXS cross section for spin excitations (i.e. the magnetic part of the spectrum extending up to ca. 0.8 eV) is directly proportional to the two-point dynamical spin-spin correlation function (local RIXS form factors are independent on transferred momentum here), see also Refs. 9 and 11-14. Therefore, the magnetic part of the RIXS spectrum was fitted to the two-point dynamical spin-spin correlation function (see next paragraph), convoluted with a gaussian to account for experimental resolution plus a low-energy gaussian covering the elastic and phonon peak around 50 meV. This was capturing most of the measured intensity with nearly constant *J*, amplitude and resolution. Fixing these 3 parameters to their mean value gives reasonable fits for all ***q***. The time-normalized data sets were corrected for the effective scattering volume $V_{eff}$

$$\frac{1}{V_{eff}} = 1 + \frac{\mu_1}{\mu_2}(\omega) \cdot \frac{\sin\theta}{\sin\beta}(q)$$

where $(\sin\theta/\sin\beta)(\boldsymbol{q})$ is a geometric term, θ (β) being the angle between the sample surface and the direction of photon incidence (detection) and $(\mu_1/\mu_2)(\omega)$ an outgoing-energy dependent absorption part derived from the total fluorescence yield (TFY) X-ray absorption spectra.

The exact solvability of the Heisenberg spin-½ chain is exploited to compute its two-point dynamical spin-spin correlation function. At zero magnetization, this is predominantly carried by two- and four-spinon intermediate states whose exact contributions can be written in the thermodynamic limit in terms of fundamental integrals by using the Vertex Operator Approach, cf. Ref. 22. The numerical evaluation of these integrals provides the dynamical structure factor throughout the BZ, yielding sum rule saturations of the order of 98%.

### *C. Theory: orbital excitations*

We first calculate the energies of the 3*d* orbitals in $Sr_2CuO_3$ using the *ab-initio* quantum chemistry method [see Sec. 5 of supplementary information]. While this method is not tailored at calculating the dispersion of orbital excitations (for which model calculations are needed) due to the finite size of the cluster, it identifies from first principles the energies of orbital excitations in $Sr_2CuO_3$ for $q$=0 transferred momentum, which have been found to be in close agreement with both optical absorption and RIXS experiments (see Ref. 19). Owing to the large separation of the *d-d* excitations energies [see Table I in Sec. 5 of supplementary information] we can separately model the momentum dependent RIXS cross section for each orbital excitation for the case of non-negligible inter-orbital hopping. For this (sec Sec. C1 below) the Kugel-Khomskii Hamiltonian for the case of a single orbital excitation (e.g. the $x^2$-$y^2$ to *xz* transition) from a ferro-orbital ground state (i.e. all $x^2$-$y^2$ orbitals are occupied) is derived based on the established charge transfer model for $Sr_2CuO_3$. Similar calculations are performed for the other orbital excitation symmetries (see also Sec. C1 below) and the RIXS cross section based on the solution of these model calculations is calculated (see section C2 below).

## *C1. Model Hamiltonian*

In Mott insulators the Kugel-Khomskii model is the effective low-energy superexchange model for coupled spin and orbital excitations, cf. Ref. 1. We consider first only an *xz* orbital excitation, which may hop between the copper sites via a "three-step" perturbative superexchange process with an energy scale $\sim t_1 \cdot t_2 /(2U)$. First (i) the particle in the excited orbital moves from site *j* to the neighbouring site *j*+1 by hopping $\sim t_2$, next (ii) an intermediate state, in which two particles are on the same site *j*+1 and which costs the Coulomb repulsion $\sim U$, is formed and finally (iii) the particle in the ground state orbital moves from site *j*+1 to site *j* by hopping $t_1$. Thus, one obtains

$$H_O = -J_O \sum_{j,s}(c^+_{j\sigma}c_{j+1,\sigma} + h.c.) + E_o \sum_j (1-n_j) \quad \text{(1S)},$$

where $J_O = (3R_1 + R_2) 2t_1 \cdot t_2 / U$ and $E_o$ is the energy cost of the orbital excitation, while $R_1=1/(1-3\eta)$ and $R_2=1/(1-\eta)$ originate from the multiplet structure of the intermediate states of the superexchange processes and depend on the ratio $\eta = J_H / U$ of the Hund's exchange to the Coulomb repulsion. While the above equation is given for the Mott-Hubbard limit of the superexchange model, in our calculation we modified the parameters to account for superexchange processes on oxygen atoms in $Sr_2CuO_3$. For a standard set of charge transfer parameters of Ref. 24 this gives $J_O \sim 0.075$ eV without any fitting to the experimental data. The values of $E_o$ are determined from the RIXS experiment and are in 10% accuracy with the on-site orbital energies obtained using *ab initio* quantum chemistry cluster calculations[19] for four $CuO_4$ plaquettes in $Sr_2CuO_3$, see also Sec. 5 of the supplementary information.

In (1S) $c_{js}$ is a fermion operator acting in the restricted Hilbert space with no double occupancies which *creates* an orbital excitation (hole in the spin background in the "t-J model language") at site $j$ with spin $s$, while $1-n_j = (1-\sum_s c^+_{j\sigma}c_{j\sigma})$ is the number operator which counts the number of orbital excitations in the chain. Note that this "fermionization" of the 1D problem was performed by replacing the standard orbital pseudospin operators[23] by fermions via Jordan-Wigner transformation, see Ref. 8 for details.

The second part of the Hamiltonian in equation (1) corresponds to the spin dynamics on the bonds where the orbital excitation is not present:

$$H_S = J \sum_j S_j S_{j+1} \quad (2S)$$

where $J$ ( $= 4t_1^2/U$ ) is now the well-known Heisenberg AF superexchange constant. $H_S$ and $H_O$ constitute the Hamiltonian $H$ from Eq. 1 in the main text.

Excitations from the ground state $3d\ x^2-y^2$ orbital to either $xy$ or $xz$ orbitals are both described by $H$, but with different effective parameters since the hopping from one of these orbitals to the neighbouring bonding oxygen orbitals and subsequently to the neighbouring copper sites differ – in particular for the $xy$ orbital the effective hopping along the chain is ca. 25% smaller than for the $xz$ orbital due to the formation of bonding and antibonding states with $2p$ orbitals on oxygens outside the Cu-O chain. For orbital excitations involving the $yz$ and $3z^2-r^2$ orbitals the dispersion vanishes, since the hopping matrix elements to the neighbouring oxygen orbitals along the chain are either much smaller than for the $xy$ or $xz$ orbital excitations ($3z^2-r^2$ orbital) or even vanishingly small ($yz$ orbital), cf. Ref. 24.

## *C2. RIXS cross section*

Local effective RIXS operators were derived for the $Cu^{2+}$ ($3d^9$) electronic configuration by Luo et al.[12] and Van Veenendaal[30] (see also Sec. 3 of supplementary information). However, several simplifications beyond the local effective scattering operators as used in Refs 13,28 are justified, because the spin of the particle in the excited orbital is to good approximation conserved (Hund's exchange is one order of magnitude smaller than Coulomb repulsion[23-24] during orbiton propagation). First, the spectrum obtained is independent of the spin of the orbital excited. Secondly, the spin and the orbital character of the excited orbital are conserved during propagation. The RIXS intensity for an orbital excitation is, therefore, given by the sum of the intensities starting from an up and down spin with either the spin flipped or conserved in the final state. This allows expressing the RIXS intensity as the product[11,13] of polarization-dependent intensities and the non-local dynamical structure factor $O(k,\omega)$ defined as the spectral function of the single orbital excitation created using operator $c_{js}$ and propagating with Hamiltonian $H$, cf. Ref. 8. The orbital dynamical structure factor $O(k, \omega)$ is either equal to unity for "non-dispersive" orbital excitations ($3z^2$-$r^2$ and $yz$ orbitals, see Sec. C1) or is calculated numerically by exactly solving the Hamiltonian $H$ on a finite chain of 28 sites. The latter is done separately for $xz$ and $xy$ orbital excitations; note that such a separation is possible as orbiton-orbiton interactions vanish for one-orbiton excitations in the chain. The results of these calculations are presented in Fig. 4c of the main text.

# Supplementary information

# Spin-Orbital Separation in the Quasi 1D Mott-insulator $Sr_2CuO_3$

J. Schlappa[1,2*], K. Wohlfeld[3], K. J. Zhou[1], M. Mourigal[4], M. W. Haverkort[5], V. N. Strocov[1], L. Hozoi[3], C. Monney[1], S. Nishimoto[3], S. Singh[6#], A. Revcolevschi[6], J.-S. Caux[7], L. Patthey[1], H. M. Rønnow[4], J. van den Brink[3], and T. Schmitt[1*]

[1]Paul Scherrer Institut, Swiss Light Source, CH-5232 Villigen PSI, Switzerland

[2]Institut Methoden und Instrumentierung der Forschung mit Synchrotronstrahlung G-I2, Helmholtz-Zentrum Berlin für Materialien und Energie GmbH, D-12489 Berlin, Germany

[3]Institute for Theoretical Solid State Physics, Helmholtzstrasse 20, 01069 Dresden, Germany

[4]Ecole Polytechnique Fédérale de Lausanne (EPFL), CH-1015 Lausanne, Switzerland

[5]Max Planck Institute for Solid State Research, D-70569 Stuttgart, Germany

[6]ICMMO - UMR 8182 - Bât. 410, Université Paris-Sud 11, 91405 Orsay Cedex, France

[7]Institute for Theoretical Physics, University of Amsterdam, Science Park 904, Postbus 94485, 1090 GL Amsterdam, The Netherlands

Present address:

[#]Indian Institute of Science Education and Research, 900 NCL Innovation Park, Pashan – 411008 Pune, India

[*]**Corresponding authors:**

**justine.schlappa@helmholtz-berlin.de & thorsten.schmitt@psi.ch**

**1. Sr$_2$CuO$_3$ compound**

Sr$_2$CuO$_3$ is built up of CuO$_4$ plaquettes, arranged in chains with two plaquettes sharing a corner oxygen atom (fig. S1). The central atom of each plaquette contains a Cu ion in $3d^9$ configuration, corresponding to a hole (called 'particle' throughout the paper, see Sec. 2) in the 3$d$ shell that carries a spin s=½. Due to the large on-site Coulomb repulsion $U$ between the Cu 3$d$ electrons, a Mott-insulating state arises. The optical gap of about 1.5 eV is of charge-transfer type, i.e. due to transfer of O 2$p$ electrons to the Cu 3$d$ shell. The large crystal-field anisotropy of the quasi 1D system leads to a ferro-orbital ground state, where the particles occupy on each plaquette Cu 3$d$ $x^2$-$y^2$ orbitals. The AF superexchange coupling $J$ between spins on neighbouring plaquettes is with $J$~250 meV exceptionally large[19,22] giving Sr$_2$CuO$_3$ the nearly ideal properties of a 1D AF spin-½ Heisenberg chain.

**2. Particle-hole transformation**

As mentioned in Sec. 1 above, the Cu$^{2+}$ ion in Sr$_2$CuO$_3$ has 9 electrons in the 3$d$ shell. Therefore, it is useful to make an electron-hole transformation and to discuss the physics of this compound in the 'hole' language: thus, as already pointed in Sec. 1, in the ground state there is only one hole in the 3$d$ $x^2$-$y^2$ orbital so that, e.g., 'making an $xy$ orbital excitation' corresponds to moving a hole from the $x^2$-$y^2$ to the $xy$ orbital. However, we refer to these holes as 'particles' in the article, in order to avoid confusion with 'holes' defined throughout the paper as empty sites in the AF chain (which are in fact made up of two holes in the 'hole' language).

We implicitly use the 'hole' language to describe $Sr_2CuO_3$ in the paper, except for the four introductory paragraphs of the main text [including Fig. 1a-b] and for Sec. 3 with fig. S3 in this supplementary information (where in both cases it is much more customary to use the electron picture).

## 3. RIXS scattering process for orbital excitations

Resonant inelastic X-ray scattering (RIXS) is a photon-in photon-out technique for which choosing the appropriate incident photon energy to the absorption resonance tremendously enhances the cross section for specific excitations[9]. The measured photon energy transfer $\Delta E = h\nu_{in} - h\nu_{out}$ and momentum transfer $q = k_{in} - k_{out}$ in a RIXS experiment is directly related to the energy and momentum of the created excitations (spinons, orbitons). For soft X-rays around 930 eV it is possible to probe $q$ of the order of the size of a Brillouin Zone of $Sr_2CuO_3$.

An orbital excitation in the RIXS scattering process is illustrated in fig S3[†]. When the photon energy is tuned to the Cu $L_3$ edge, in the intermediate state the $3d$ $x^2-y^2$ orbital is occupied with two electrons, creating at the same time a $2p$ core hole in the $j=3/2$ state. In the following decay process the core hole can be refilled from a $3d$ orbital with a different symmetry, e.g. $xz$. Each transition has to obey the electric dipole (E1) selection rules, leading to specific polarization dependencies. In general, transitions of an electron are allowed from, e.g., $xz$ to the $x^2-y^2$ orbital, either with spin up, spin down, or flipping an up (down) spin to down (up), each with a different resonant enhancement and polarization dependence. Local effective RIXS operators were derived for Cu $d^9$ by Luo et al.[12] and Van Veenendaal[30].

---

[†] Unlike in the rest of the paper, where the 'hole' picture is used, in this section we use the 'electron' picture to stay in agreement with Ref. 9.

**4. Further experimental details**

The set up of our experiment is depicted in fig. S1 (for description see Methods summary and Methods). The variation of the projection of the incident-photons' polarization vector $\boldsymbol{\varepsilon}_{in}$ in the [100] and [010] directions, resulting from sample rotation for $q$-dependent measurements, is shown in fig. S2b. The momentum transfer along the chains, $\boldsymbol{q}_a$, is denoted with $q$ in the main text.

**5. Calculating the energies of Cu 3*d* orbitals using quantum chemistry *ab-initio* calculations**

To acquire better insight into the Cu 3$d$-level electronic structure of $Sr_2CuO_3$ and the nature of the orbital excitations quantum chemical calculations were performed. In the spirit of modern multiscale electronic-structure approaches, we describe a given region around a central Cu site by advanced *ab initio* many-body techniques, while the remaining part of the solid is modeled at the Hartree-Fock level. The complete-active-space self-consistent-field (CASSCF) method was used to generate multireference wavefunctions for further configuration-interaction (CI) calculations, see e.g. Ref. 31. In the CASSCF scheme a full CI is carried out within a limited set of "active" orbitals, i.e., all possible occupations are allowed for those active orbitals. The active orbital set includes here all 3$d$ functions at the central Cu site and the 3$d$ $x^2$-$y^2$ functions of four Cu nearest neighbours. Strong correlations among the 3$d$ electrons are thus accurately described. The multireference CI (MRCI) calculations incorporate on top of the CASSCF wavefunction all single and double excitations from the Cu 3$s$, 3$p$, 3$d$ and O 2$p$ orbitals on a given CuO$_4$ plaquette and from the 3$d$ $x^2$-$y^2$ orbitals of the four Cu nearest neighbours. To the MRCI energies we also added Davidson corrections that account for quadruple excitations in the CI expansion, see Ref. 31. Further technical details are discussed in Methods and in Ref. 19. It turns out by MRCI that the lowest crystal-field excitation at 1.62 eV is to the $xy$ level. Excitations to the $xz$ and $yz$ orbitals require 2.25 and 2.32 eV, respectively, while the splitting between the $x^2$-$y^2$ and 3$z^2$-$r^2$ levels is 2.66 eV. These values were computed for a ferromagnetic cluster, i.e., they do not include the effect of superexchange interactions in Sr$_2$CuO$_3$ (which would shift the energies by ca. ~ $J$). The results are tabulated in Table 1. Finally, let us note that we obtain a quantitative similar result using LDA Wannier orbitals[32,33] combined with multiplet ligand field theory.

|  | $xy$ | $xz/yz$ | $3z^2$-$r^2$ |
|---|---|---|---|
| Experiment | 1.85 | 2.36 | 2.98 |
| Theory | 1.62 | 2.25/2.32 | 2.66 |

**Table 1 | Comparison between experimental and theoretical values (quantum chemistry cluster calculations) of on-site orbital excitations.**

**6. Ruling out competing scenarios to the $J_O$-$J$ –model**

The structure factor $O(\mathbf{k},\omega)$ (see Methods) reproduces correctly the experimental results. If one assumes that $O(\mathbf{k},\omega)$ is equal to unity also for the "dispersive" orbital excitations ($xz$ and $xy$), then the result (fig. S4, left) cannot properly reproduce the experimental RIXS cross section. In other words, the observed orbital dispersion cannot be attributed to a mere shift of spectral weight. In fact, this single ion picture[9-10] does not even correctly predict the intensity of the excitations: for example for the *xy* orbital excitation it suggests that it does not depend on momentum and thus it can only be the spin-orbital separation mechanism transferring the weight from the orbiton to the spinon-orbiton continuum.

Also a non-interacting band picture cannot explain the $\pi$ periodicity of the spectrum, nor can it explain a large incoherent spectrum. We have also calculated the spectral function $O(\mathbf{k},\omega)$ using the mean-field decoupling of spin and orbitals [see e.g. Ref. 23] and then by calculating the orbiton dispersion similarly as in linear spin wave theory; in such a mean-field approach the hopping of the orbital excitation is by definition (and artificially) decoupled from the spin excitation. The impact of the spin background is in this case only reflected in a renormalization of $J_O$. Therefore, spin-orbital separation is absent and the spectral function of a particular orbital excitation (*xy* or *xz*) consists of a single dispersive quasiparticle peak with period $2\pi$. As shown in the right panel of fig. S4, such results do not reproduce the experiment and indeed the exact diagonalization of the Hamiltonian *H* is needed to correctly reproduce the experimental results and in particular the observed $\pi$ periodicity. Moreover, the particular shape of the RIXS cross section is due to the spin-orbital separation which cannot be captured by a mean-field approach.

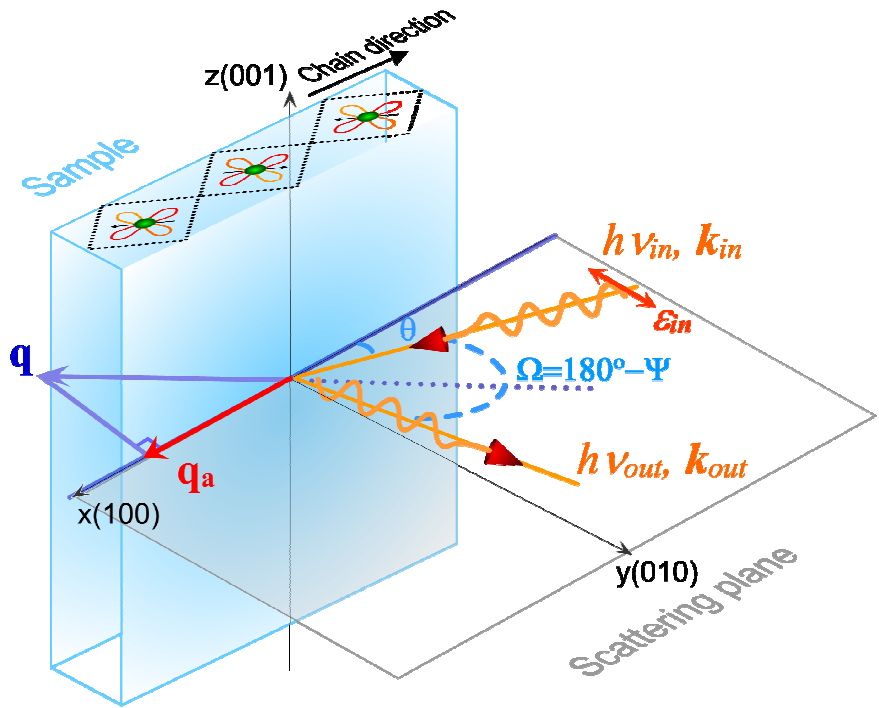

**Figure S1 | Experimental geometry.** Experimental geometry in our RIXS set up, see text for details. The orientation of the chains in $Sr_2CuO_3$ – built up from corner-sharing $CuO_4$ plaquettes (see also Fig. 1d main text) is shown: the green dots indicate the position of the Cu-sites with the lobes representing $3d\ x^2\text{-}y^2$ orbitals.

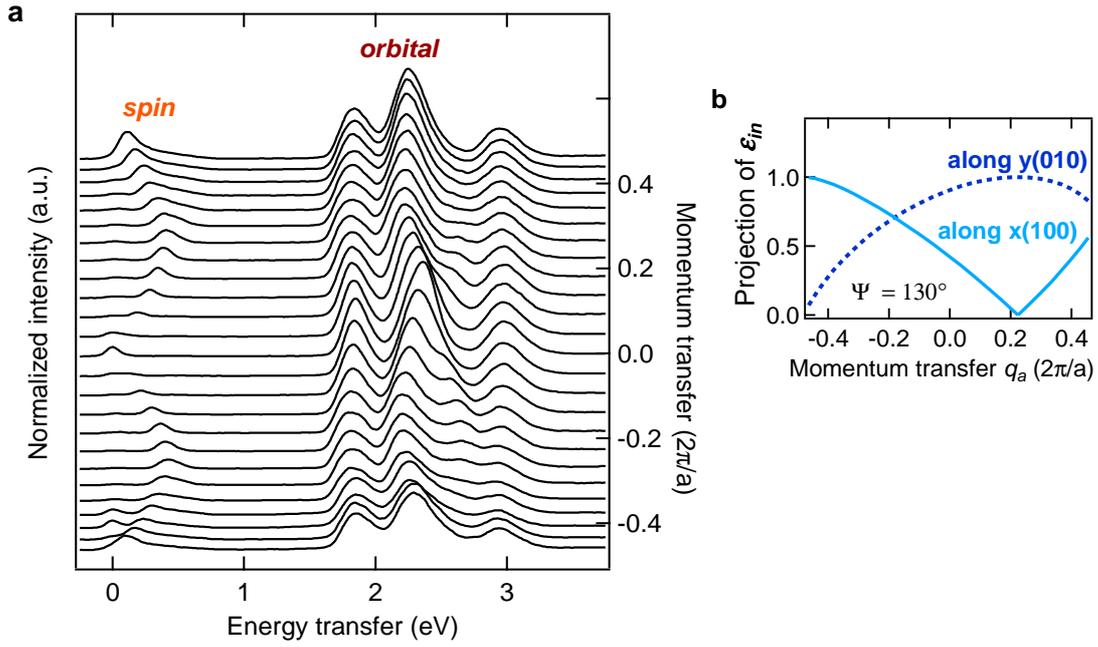

**Figure S2 | RIXS line spectra and polarization dependence. a,** RIXS line spectra for Ψ=130° (the same data set as the intensity map in Fig. 1c, main text). **b,** change of projection of the polarization vector of incoming photons, $\varepsilon_{in}$, along the crystallographic directions (100) and (010), which is related to rotation of the sample for momentum-dependent measurements (Ψ=130°). See text for details.

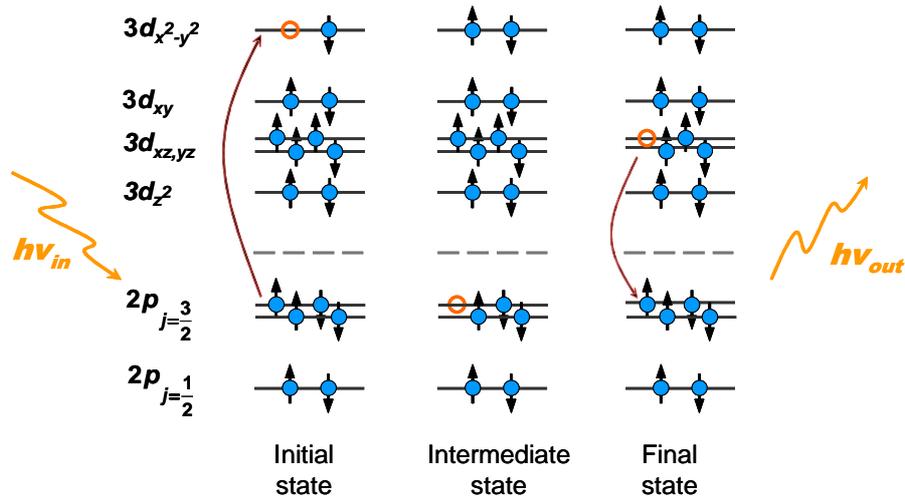

**Figure S3 | RIXS scattering process for orbital excitations.** Creation of an *xz* orbital excitation on a $Cu^{2+}$ ion. Note that here for clarity the 'electron' picture is used – in contrast to the rest of the paper where the 'hole' picture is used.

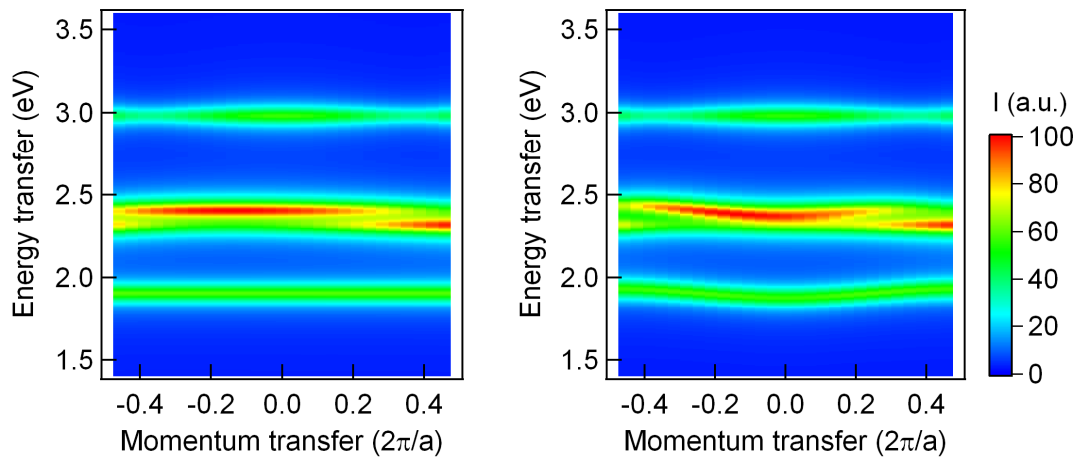

**Figure S4 | Ab-initio calculations: alternative scenarios to the spin-orbital separation process.** Simulation of the experimental data with the single-ion cross section (left) and with the linear orbital wave theory (right).